\documentclass{iau_FM}
\usepackage{graphicx}

\usepackage{natbib}
\usepackage{amssymb,amsmath} 
\newcommand{\teff}{\mbox{$T_{\rm eff}$}}

\newcommand{\rsun}{\mbox{$R_{\ensuremath{\odot}}$}}
\newcommand{\kep}{\mbox{\textit{Kepler}}}

\title[Asteroseismology of Exoplanet Host Stars] %% give here short title %%
{Asteroseismology of Exoplanet Host Stars}

\author[Daniel Huber]   %% give here short author list %%
{Daniel Huber$^{1,2,3}$
}

\affiliation{$^1$Sydney Institute for Astronomy (SIfA), School of Physics, University 
of Sydney, NSW 2006, Australia; email: {\tt daniel.huber@sydney.edu.au} \\[\affilskip]
$^2$SETI Institute, 189 Bernardo Avenue, Mountain View, CA 94043, USA \\ 
$^3$Stellar Astrophysics Centre, Department of Physics and Astronomy, Aarhus University, 
Ny Munkegade 120, DK-8000 Aarhus C, Denmark}

\pubyear{2015}
\jname{Astronomy in Focus, Volume 1} 
\editors{Piero Benvenuti, ed.}
\begin{document}

\maketitle

\begin{abstract}
Asteroseismology is among the most powerful observational tools to determine fundamental 
properties of stars. Space-based photometry has recently enabled the systematic detection 
of oscillations in exoplanet host stars, allowing a combination of asteroseismology with 
transit and radial-velocity measurements to characterize planetary systems. 
In this contribution I will review the key synergies between asteroseismology and 
exoplanet science such as the precise determination of radii 
and ages of exoplanet host stars, as well as applications of asteroseismology to 
measure spin-orbit inclinations in multiplanet systems and orbital eccentricities of 
small planets. Finally I will give a brief outlook on asteroseismic studies of 
exoplanet hosts with 
current and future space-based missions such as K2 and TESS.
\end{abstract}

\firstsection % if your document starts with a section,
\section{Introduction}

Exoplanet science has undergone a revolution in the 
past decade driven by high precision photometry from space-based missions such as CoRoT 
and \textit{Kepler}. At the time of writing of this review \textit{Kepler} has detected more 
than 4000 planet candidates \citep[e.g.][]{mullally15}, with breakthrough 
discoveries including measurements of planet densities in multi-planet 
systems \citep[e.g.][]{lissauer11} and the detection of 
planets in the habitable zone \citep[e.g.][]{quintana14}. 
In parallel, \textit{Kepler} has enabled 
tremendous observational progress in asteroseismology, the study of stellar 
pulsations.
In particular, the number of cool stars with detected oscillations driven by surface convection 
(solar-like oscillations) has increased by more than a order of magnitude, 
allowing unprecedented studies of their fundamental 
properties and interior structure \citep[see][for a recent review]{chaplin13d}.

\begin{figure}
% \vspace*{-2.0 cm}
\begin{center}
\resizebox{\hsize}{!}{\includegraphics{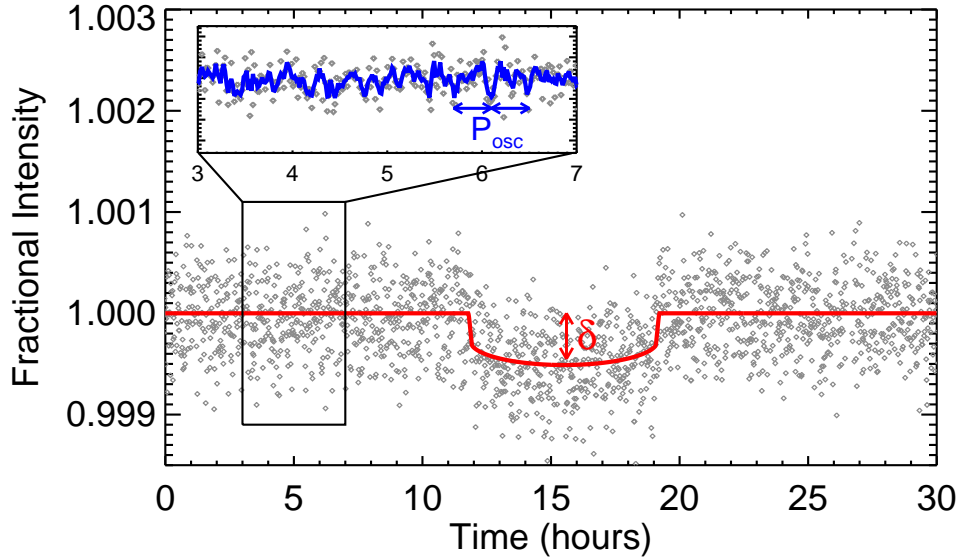}}
% \vspace*{-1.0 cm}
 \caption{\textit{Kepler} short-cadence data showing a single transit of 
Kepler-36c \citep{carter12}. 
The red solid line is the transit model, and 
the inset shows the oscillations of the host star.
The transit depth $\delta$ yields the size of the planet relative to the 
star, and the oscillation periods ($P_{\rm osc}$) can be used to 
independently measure the size of the star.}
   \label{fig1}
\end{center}
\end{figure}

The need for continuous high-precision time domain data (either in intensity or velocity) 
has enabled a fortuitous synergy between asteroseismology 
and exoplanet science since the data can be simultaneously used to detect exoplanets 
and study stellar oscillations (Figure \ref{fig1}). In this review I will discuss  
recent key synergies between asteroseismology and exoplanet science, 
and conclude with an 
outlook of what can expected from current and future space-based missions such as K2 and 
TESS (with a particular focus on asteroseismology of evolved host stars).

\section{Precise Characterization of Exoplanets}

Indirect detection methods such as transit and radial velocity surveys
measure relative properties of exoplanets, 
hence requiring precise characterizations of their host stars. 
First asteroseismic studies of exoplanet host stars 
were performed using ground-based radial-velocity data \citep{bouchy05,bazot05} or 
space-based photometry using the Hubble Space Telescope \citep{gilliland11} and 
CoRoT \citep{ballot11b,lebreton14}.
Following early asteroseismic studies supporting \textit{Kepler} exoplanet 
detections \citep{cd10,batalha10,howell11}, \citet{huber13} presented the first 
systematic study for 77 \textit{Kepler} host stars. In addition to the precise determination of 
a large number exoplanet radii, the results yielded systematic biases for
surface gravities (and hence radii) for evolved 
stars based on high-resolution spectroscopy. The \textit{Kepler} asteroseismic 
host star sample has since been used to calibrate more indirect methods to 
determine fundamental properties of exoplanet host stars such as high-resolution 
spectroscopy \citep{brewer15} and stellar granulation \citep{bastien13}.

\begin{figure}
\begin{center}
\resizebox{12cm}{!}{\includegraphics{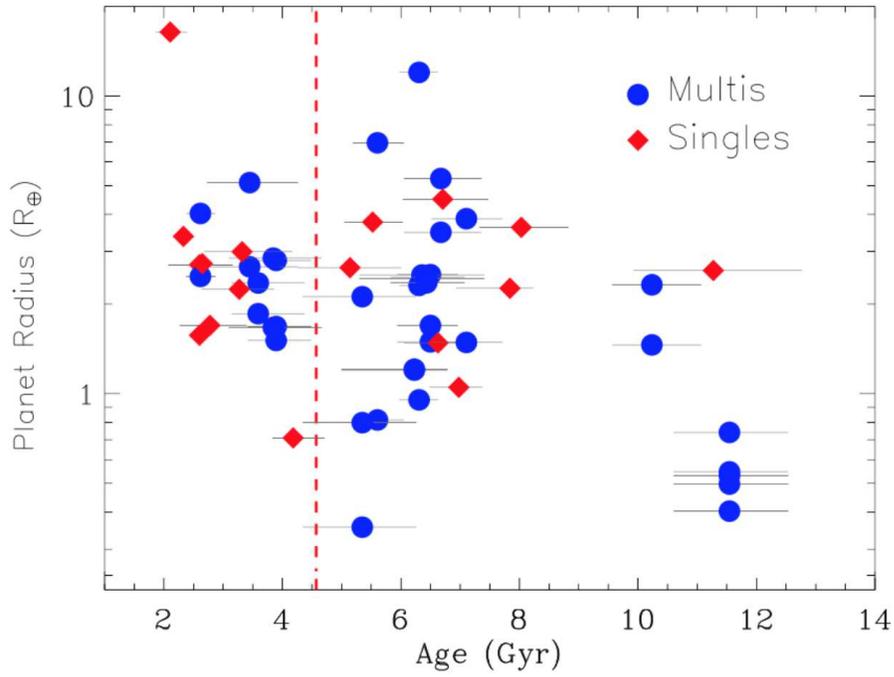}}
\caption{Planet radius versus host star age for a sample of 33 \textit{Kepler} exoplanet 
host stars. The vertical dashed line marks the age of the Sun. Note that the bimodal 
distribution is mostly due to an asteroseismic detection bias towards more luminous 
stars (see text). Adapted from \citet{silva15}.}
\label{fig2}
\end{center}
\end{figure}

More recent studies have focused on not only measuring global asteroseismic quantities 
(which are sensitive to densities, masses and radii) but also systematic 
modeling of individual oscillation frequencies, which allows precise constraints on the 
stellar age \citep[e.g.][]{metcalfe12}. Figure 
\ref{fig1} shows the distribution of radii and ages of \textit{Kepler} host stars from the 
first systematic age study by \citet{silva15}, based on individual frequencies measured 
by \citet{davies15}. While the bimodal age distribution is mostly 
due to a detection bias towards evolved subgiants ($\sim$\,6\,Gyrs) and 
more luminous F stars ($\sim$\,2-3\,Gyrs), the analysis illustrated the exquisite precision 
with which ages can be determined ($\sim$\,14\%). One of the most remarkable discoveries 
in the sample is Kepler-444, which consists of a $11.2\pm1.0$\,Gyr year old K dwarf hosting
five sub-Earth planets with orbital periods of less than 10 days \citep{campante15}. The 
system demonstrated that sub-Earth sized planets have existed for most of the 
history of our Universe, in line with earlier findings that the formation of small
planets in independent of host star metallicity \citep{buchhave12}.

\section{Spin-Orbit Inclinations and Architectures of Exoplanet Systems}

In addition to fundamental properties, asteroseismology can be used 
to determine the line-of-sight inclination of the stellar 
rotation axis by measuring relative heights of rotationally split modes \citep{gizon03}. 
For transiting exoplanets, a low stellar inclination automatically yields a 
misalignment of the orbital plane and the stellar equatorial plane 
(a high obliquity), while an inclination near 90 degrees implies that the star and the 
planets are likely (but not necessarily) aligned.

The long continuous time series by \textit{Kepler} and CoRoT allowed first 
asteroseismic stellar inclination measurements for transiting exoplanet systems 
\citep{chaplin13c,gizon13,vaneylen14}. An 
intriguing example is Kepler-56, a red giant hosting two transiting planets 
confirmed through transit-timing variations \citep{steffen12}. The Kepler-56 
power spectrum revealed an 
inclination of $47\pm6$ degrees, 
demonstrating the first stellar spin-orbit misalignment in a multiplanet 
system \citep{huber13b}.
Recent studies have since also determined the spin-axis inclination 
in systems for which the projected obliquity has been measured using the 
Rossiter-McLaughlin effect, which allows a 
determination of the 3D obliquity angle \citep{lund14,benomar14}. 

\begin{figure}
\begin{center}
\resizebox{\hsize}{!}{\includegraphics{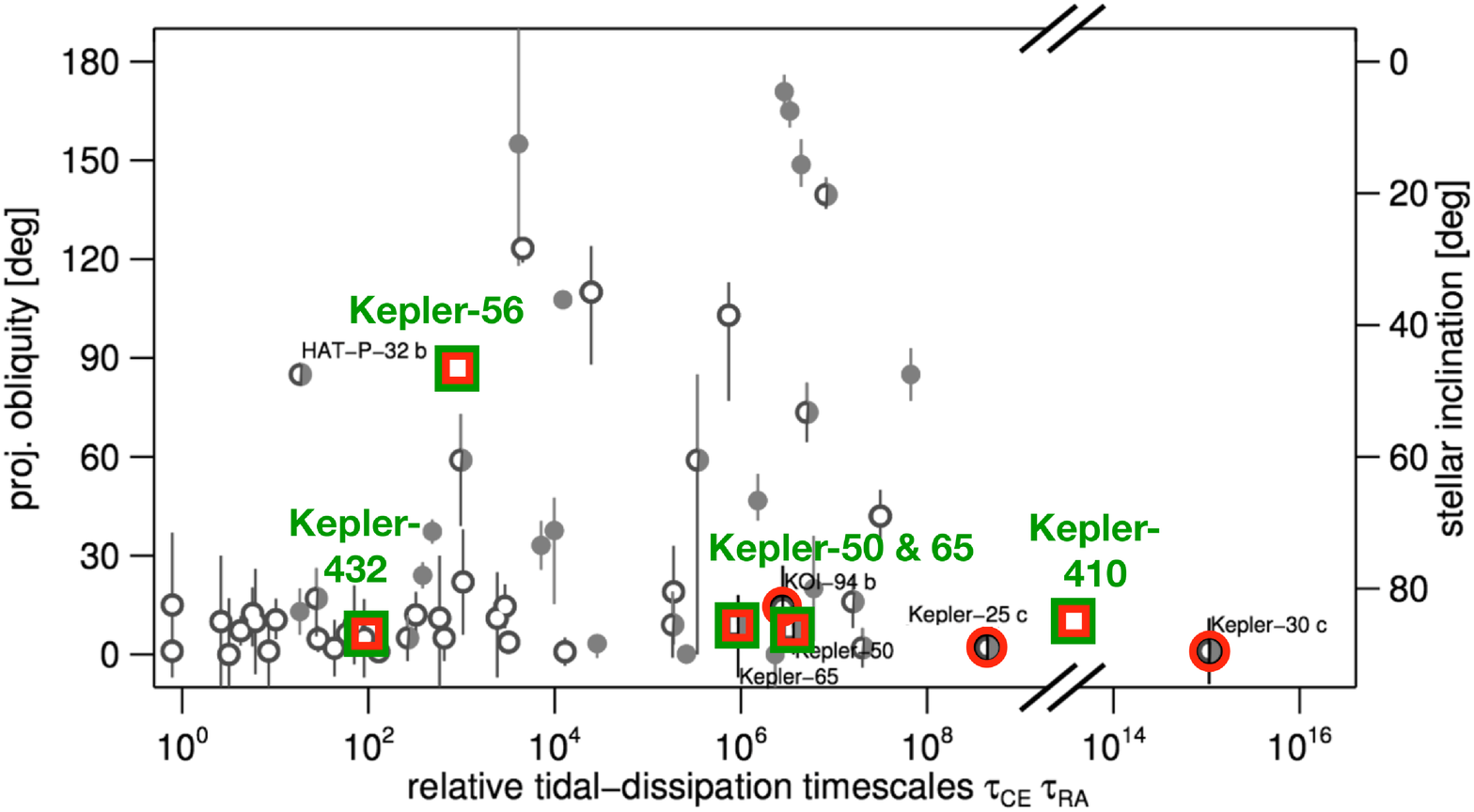}}
\caption{Projected obliquity (left ordinate) and stellar inclination (right ordinate) 
versus relative tidal 
dissipation timescales for exoplanet systems. Systems with short dissipation timescales 
are expected to have been re-aligned even if they were misaligned by the formation 
process, while systems with long dissipation timescales are expected to preserve their 
configuration. Multiplanet systems without hot Jupiters are highlighted 
by red circles, and 
systems with inclinations measured using asteroseismology are highlighted with green 
squares. Positions for Kepler-45, Kepler-410 and Kepler-432 are 
approximate only. Adapted from \citet{albrecht13}.
}
\label{fig3}
\end{center}
\end{figure}

Asteroseismic inclination measurements have played an increasingly 
important role for constraining formation theories for hot Jupiters.
Figure \ref{fig3} shows the projected obliquity or stellar 
inclination for exoplanet systems as a function 
of the relative tidal dissipation timescale, a proxy for how quickly a system 
can be realigned by tidal interactions if it was initially misaligned by the formation 
process \citep{albrecht12}. Hot Jupiter systems with intermediate 
dissipation timescale are frequently observed to have high obliquities 
\citep[e.g.][]{winn10}, while coplanar multiplanet systems without hot Jupiters 
have have mostly low 
obliquities \citep[e.g.][]{sanchis13} despite long tidal dissipation timescales. 
This has been taken as evidence that the formation of hot Jupiters is related to 
dynamical interactions, rather than migrations through a protoplanetary disk.

The misalignment in the Kepler-56 system has yielded the first outlier in this trend. 
However, other seismic measurements yielded well aligned multiplanet 
systems, including the recent discovery 
of the Kepler-432 system which consists of a red giant with a transiting 
warm Jupiter and an outer giant planet detected through radial velocities 
\citep{quinn15}. Remarkably, Figure \ref{fig3} shows that 
most constraints on orbital architectures for 
small multiplanet systems come from asteroseismology, since the technique is 
independent of planet size (unlike Rossiter-McLaughlin observations). 
Future asteroseismic inclination measurements will increase 
the sample and help to determine whether  
spin-orbit misalignments in multiplanet systems without hot Jupiters are common.

\section{Orbital Eccentricities using Asteroseismic Densities}

Exoplanet transits allow a measurement of the 
semi-major axis relative to the stellar radius ($a/R_{*}$), 
provided the eccentricity of the orbit is known. 
For circular orbits, $a/R_{*}$ is related to the mean 
density of the star \citep[see, e.g.,][]{seager03}:

\begin{equation}
\rho_{\star, \rm{transit}} = \frac{3\pi}{GP^{2}}\left(\frac{a}{R_{\star}}\right)^3 \: ,
\label{equ:rho}
\end{equation}

\noindent
where $G$ is the gravitational constant and $P$ is the orbital period. 
For non-circular orbits, it can be shown that the 
true stellar density ($\rho_{\star}$) is related to the density 
measured from the transit assuming a circular orbit ($\rho_{\star, \rm{transit}}$) 
as \citep[e.g.][]{kipping10}:

\begin{equation}
\frac{\rho_{\star}}{\rho_{\star, \rm{transit}}} = \frac{(1-e^{2})^{3/2}}{(1+e \sin{\omega})^{3}}
\end{equation}

\noindent
where $e$ is the 
orbital eccentricity, and $\omega$ is the argument of periastron. Equations (1) and (2) 
demonstrate that if an independent measurement of $\rho_{\star}$ is available, 
transits can be used to directly constrain the eccentricity of a planet without the 
need for radial velocity observations. Incidentally, a 
key observable in asteroseismology is the sound travel time through the 
stellar diameter, which for an ideal gas is directly related to the square root of the mean 
stellar density \citep{kb95}. Comparisons of asteroseismic densities with densities 
derived from transiting exoplanets with known 
eccentricities \citep{gilliland11,nutzman11,huber14b} and 
stellar models \citep{stello09,white11} have shown excellent 
agreement for solar-type stars at the level of a few percent. 

\begin{figure}
\begin{center}
\resizebox{\hsize}{!}{\includegraphics{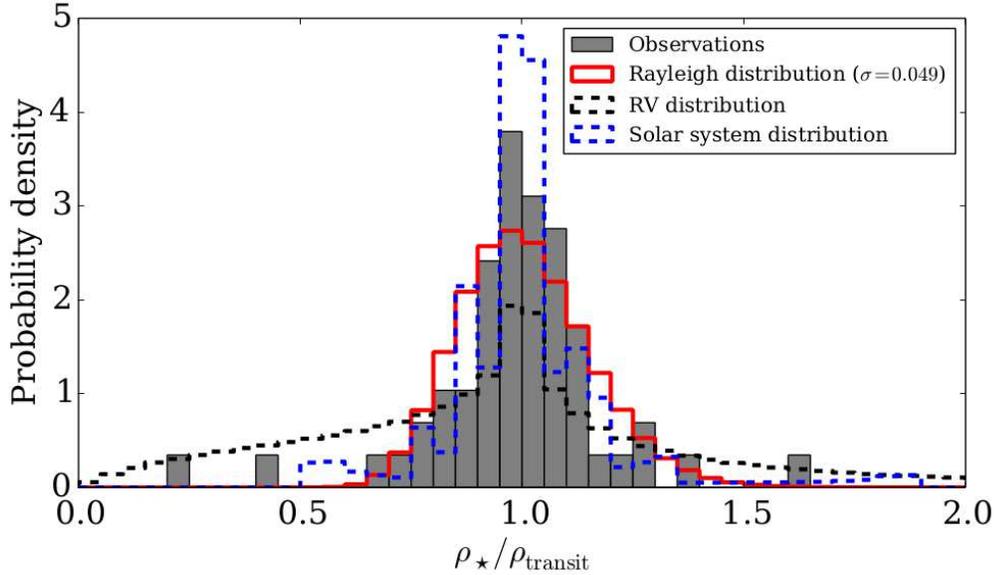}}
\caption{Ratio of the true mean stellar density and the density measured from transits 
assuming a circular orbit for 28 \textit{Kepler} multiplanet host stars (grey). Corresponding 
distributions of solar system planets and planet detected by radial velocity 
surveys are shown as blue and black dashed histograms, respectively. 
From \citet{vaneylen15}.}
\label{fig4}
\end{center}
\end{figure}

Following the first systematic study of eccentricities with asteroseismic 
densities by \citet{sliski14} to identify false positives among 
41 \textit{Kepler} single planet-candidate hosts, \citet{vaneylen15} concentrated on a 
sample of 28 \textit{Kepler} multiplanet systems.
Figure \ref{fig4} shows a histogram of the left hand side of equation (2) 
for their \textit{Kepler} sample compared to the solar system as well as a 
sample of planets with eccentricities determined from radial velocity surveys. 
The seismic host star sample (red solid lines) is 
consistent with circular orbits, similar to the solar system 
(blue dashed histogram) but in stark contrast to the radial velocity sample 
(black dashed line). Since the seismic host star sample includes mostly small, 
low-mass planets compared to the more massive planets probed by Doppler surveys, 
this indicates that low-mass planets are preferentially on 
circular orbits. This conclusion is of great importance since 
circular orbits are often explicitly assumed to model the habitable 
zone \citep{borucki13,barclay13,quintana14,jenkins15} or 
for estimating transit durations to account for detection completeness in 
planet occurrence studies \citep[e.g.][]{howard11,dong13b,petigura13b,burke15}. 
Furthermore, eccentricities are often a key property for constraining planet 
formation scenarios (such as for hot Jupiters, see previous Section).

\section{Future Prospects}

The repurposed \textit{Kepler}/K2 Mission is currently conducting 80 day observing campaigns in 
the ecliptic plane \citep{howell14}. 
While K2 has enabled a large amount of new science, 
it does not have the ability to follow-up transit detections made with long-cadence 
(30-minute sampling) data with short-cadence (1-minute sampling) observations 
that are required for asteroseismic analyses of solar-type and subgiant stars. 
Hence, the only way to study solar-type and subgiant host stars 
is to either target exoplanet hosts which were known before the K2 
campaign started, or revisit already observed fields.

\begin{figure}
\begin{center}
\resizebox{\hsize}{!}{\includegraphics{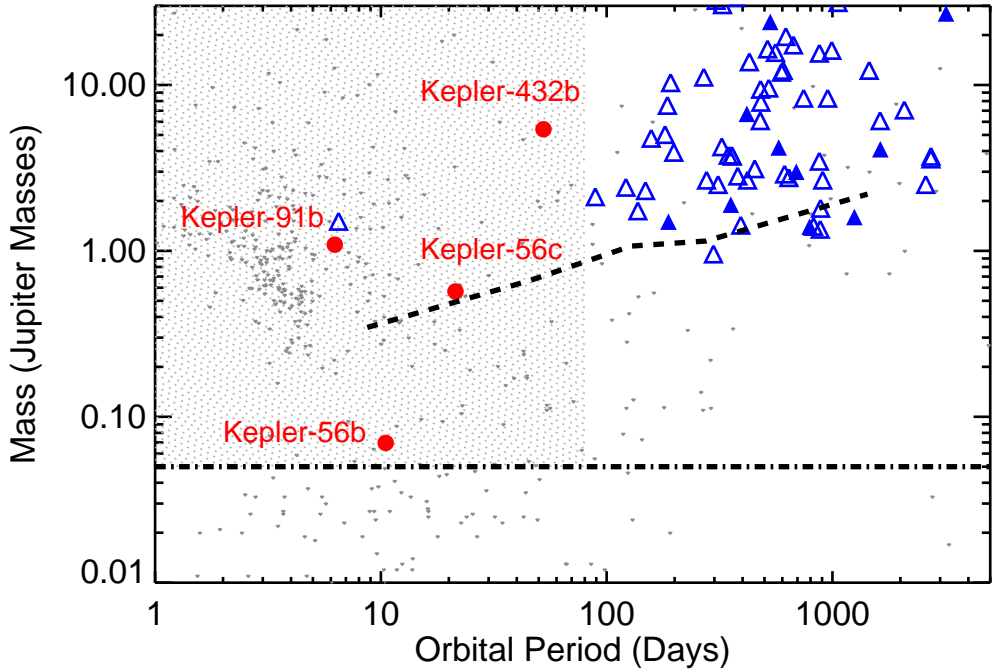}}
\caption{Mass versus orbital period for exoplanets orbiting 
red-giant branch stars ($R>3.5\rsun$, $\teff < 5500$\,K) detected with transits (red 
circles) and radial velocities (blue triangles). Open triangles are mean 
masses assuming random orbit orientations (i.e., $M \sin(i) \pi/2$) 
and grey dots show planets orbiting dwarfs.
The dashed line shows the median RV detection limit for mean masses given by \citet{bowler10}. 
The dashed-dotted line marks the mass of Neptune as an approximate K2 detection limit 
($R \gtrsim 0.5 R_{J}$), and 
the grey shaded area will be probed by K2.
Note that RV surveys are not sensitive to warm Neptune-Saturn 
mass planets which will can be detected with K2.}
\label{fig6}
\end{center}
\end{figure}

An alternative is to conduct a transit survey for
stars which are evolved enough to have oscillation periods detectable 
with long-cadence data, 
but small enough to allow the detection of planets.
Figure \ref{fig6} shows the distribution of known planets around evolved ($R>3.5\rsun$) 
stars in planet mass - orbital period plane.
While radial velocity surveys demonstrated that 
close-in planets with $>5 M_{J}$ are rare around red giants, 
\kep\ has discovered several giant planets 
orbiting oscillating RGB stars within $\lesssim$\,0.3\,AU ($P\lesssim 50$\,d) 
\citep[e.g. Kepler-91b, Kepler-56bc, Kepler432b;][]{lillobox14,huber13b,ciceri15}. 
This hints at the existence of a yet poorly studied population of
warm Neptune--Saturn mass planets around evolved stars, which has not been 
probed with radial-velocity surveys.
% targeting 
%evolved stars \citep[e.g.,][]{johnson07b,niedzielski09}, which are only sensitive to 
%more massive ($M\gtrsim 0.5 M_{J}$) planets within $\lesssim$1\,AU.

RGB stars span a wide range of stellar 
masses, luminosities, and chemical compositions, allowing studies of 
exoplanets in a variety of host star environments. Both aspects make 
RGB stars ideal targets to tackle
two controversial and unsolved questions in exoplanet science:  
the occurrence rate of gas-giant planets as a function of stellar 
mass \citep[e.g.,][]{lloyd11,johnson13,schlaufman13}, 
and the role of stellar incident 
flux on the radius inflation of gas-giant 
planets  \citep{bodenheimer01,chabrier07,lopez15}. A K2 transit survey targeting 
low-luminosity RGB stars can provide important clues to solve these problems.

In addition to K2, future missions such as the Terrestrial Exoplanet Survey Satellite 
TESS \citep{ricker14} and PLATO \citep{rauer14} will provide opportunities to perform 
asteroseismology of exoplanet host stars. While the smaller apertures of TESS compared to 
\textit{Kepler} will restrict asteroseismic studies to the brightest dwarfs and subgiants, 
planet yield simulations by 
\citet{sullivan15} indicate hundreds of 
%$\sim$\,800 
evolved host stars which may be suited for 
asteroseismic studies using 30-minute cadence full-frame observations 
(Campante et al., in prep). Additionally, the 
concept for PLATO includes the capability to detect oscillations of a large number of 
exoplanet host stars. Combined with ground-based radial-velocity efforts such as the 
SONG \citep{grundahl08} and LCOGT \citep{brown13} networks, there is little doubt that the 
synergy between asteroseismology and exoplanet science will continue to thrive over the 
coming decades.

\vspace{0.2cm}
\noindent
\textbf{Acknowledgments:}
I thank Simon Jeffery and Joyce Guzik for organizing a fantastic 
IAU focus meeting, Eric Gaidos, Andrew Howard and John Johnson for enlightening 
discussions on finding giant planets orbiting giant stars, 
as well as Victor Silva Aguirre, Simon Albrecht and 
Vincent van Eylen for providing figures and comments for this review. 
Financial support was provided by NASA grant NNX14AB92G and the Australian Research Councils 
Discovery Projects funding scheme (project number DEI40101364).

\bibliographystyle{apj}
\bibliography{/Users/daniel/Dropbox/epicpaper/tex/references.bib}

\newcommand{\SortNoop}[1]{}
\begin{thebibliography}{65}
\expandafter\ifx\csname natexlab\endcsname\relax\def\natexlab#1{#1}\fi

\bibitem[{{Albrecht} {et~al.}(2013){Albrecht}, {Winn}, {Marcy}, {Howard},
  {Isaacson}, \& {Johnson}}]{albrecht13}
{Albrecht}, S. et al. 2013, \apj, 771, 11

\bibitem[{{Albrecht} {et~al.}(2012){Albrecht}, {Winn}, {Johnson}, {Howard},
  {Marcy}, {Butler}, {Arriagada}, {Crane}, {Shectman}, {Thompson}, {Hirano},
  {Bakos}, \& {Hartman}}]{albrecht12}
{Albrecht}, S., {et~al.} 2012, \apj, 757, 18

\bibitem[{{Ballot} {et~al.}(2011){Ballot}, {Gizon}, {Samadi}, {Vauclair},
  {Benomar}, {Bruntt}, {Mosser}, {Stahn}, {Verner}, {Campante},
  {Garc{\'{\i}}a}, {Mathur}, {Salabert}, {Gaulme}, {R{\'e}gulo}, {Roxburgh},
  {Appourchaux}, {Baudin}, {Catala}, {Chaplin}, {Deheuvels}, {Michel}, {Bazot},
  {Creevey}, {Dolez}, {Elsworth}, {Sato}, {Vauclair}, {Auvergne}, \&
  {Baglin}}]{ballot11b}
{Ballot}, J., {et~al.} 2011, \aap, 530, A97

\bibitem[{{Barclay} {et~al.}(2013){Barclay}, {Burke}, {Howell}, {Rowe},
  {Huber}, {Isaacson}, {Jenkins}, {Kolbl}, {Marcy}, {Quintana}, {Still},
  {Twicken}, {Bryson}, {Borucki}, {Caldwell}, {Ciardi}, {Clarke},
  {Christiansen}, {Coughlin}, {Fischer}, {Li}, {Haas}, {Hunter}, {Lissauer},
  {Mullally}, {Sabale}, {Seader}, {Smith}, {Tenenbaum}, {Kamal Uddin}, \&
  {Thompson}}]{barclay13}
{Barclay}, T., {et~al.} 2013, \apj, 768, 101

\bibitem[{{Bastien} {et~al.}(2013){Bastien}, {Stassun}, {Basri}, \&
  {Pepper}}]{bastien13}
{Bastien}, F.~A., {Stassun}, K.~G., {Basri}, G., \& {Pepper}, J. 2013, \nat,
  500, 427

\bibitem[{{Batalha} {et~al.}(2010){Batalha}, {Borucki}, {Koch}, {Bryson},
  {Haas}, {Brown}, {Caldwell}, {Hall}, {Gilliland}, {Latham}, {Meibom}, \&
  {Monet}}]{batalha10}
{Batalha}, N.~M., {et~al.} 2010, \apjl, 713, L109

\bibitem[{{Bazot} {et~al.}(2005){Bazot}, {Vauclair}, {Bouchy}, \&
  {Santos}}]{bazot05}
{Bazot}, M., {Vauclair}, S., {Bouchy}, F., \& {Santos}, N.~C. 2005, \aap, 440,
  615

\bibitem[{{Benomar} {et~al.}(2014){Benomar}, {Masuda}, {Shibahashi}, \&
  {Suto}}]{benomar14}
{Benomar}, O., {Masuda}, K., {Shibahashi}, H., \& {Suto}, Y. 2014, \pasj

\bibitem[{{Bodenheimer} {et~al.}(2001){Bodenheimer}, {Lin}, \&
  {Mardling}}]{bodenheimer01}
{Bodenheimer}, P., {Lin}, D.~N.~C., \& {Mardling}, R.~A. 2001, \apj, 548, 466

\bibitem[{{Borucki} {et~al.}(2013){Borucki}, {Agol}, {Fressin}, {Kaltenegger},
  {Rowe}, {Isaacson}, {Fischer}, {Batalha}, {Lissauer}, {Marcy}, {Fabrycky},
  {D{\'e}sert}, {Bryson}, {Barclay}, {Bastien}, {Boss}, {Brugamyer},
  {Buchhave}, {Burke}, {Caldwell}, {Carter}, {Charbonneau}, {Crepp},
  {Christensen-Dalsgaard}, {Christiansen}, {Ciardi}, {Cochran}, {DeVore},
  {Doyle}, {Dupree}, {Endl}, {Everett}, {Ford}, {Fortney}, {Gautier}, {Geary},
  {Gould}, {Haas}, {Henze}, {Howard}, {Howell}, {Huber}, {Jenkins}, {Kjeldsen},
  {Kolbl}, {Kolodziejczak}, {Latham}, {Lee}, {Lopez}, {Mullally}, {Orosz},
  {Prsa}, {Quintana}, {Sanchis-Ojeda}, {Sasselov}, {Seader}, {Shporer},
  {Steffen}, {Still}, {Tenenbaum}, {Thompson}, {Torres}, {Twicken}, {Welsh}, \&
  {Winn}}]{borucki13}
{Borucki}, W.~J., {et~al.} 2013, Science, 340, 587

\bibitem[{{Bouchy} {et~al.}(2005){Bouchy}, {Bazot}, {Santos}, {Vauclair}, \&
  {Sosnowska}}]{bouchy05}
{Bouchy}, F., {Bazot}, M., {Santos}, N.~C., {Vauclair}, S., \& {Sosnowska}, D.
  2005, \aap, 440, 609

\bibitem[{{Bowler} {et~al.}(2010){Bowler}, {Johnson}, {Marcy}, {Henry}, {Peek},
  {Fischer}, {Clubb}, {Liu}, {Reffert}, {Schwab}, \& {Lowe}}]{bowler10}
{Bowler}, B.~P., {et~al.} 2010, \apj, 709, 396

\bibitem[{{Brewer} {et~al.}(2015){Brewer}, {Fischer}, {Basu}, {Valenti}, \&
  {Piskunov}}]{brewer15}
{Brewer}, J.~M., {Fischer}, D.~A., {Basu}, S., {Valenti}, J.~A., \& {Piskunov},
  N. 2015, \apj, 805, 126

\bibitem[{{Brown} {et~al.}(2013){Brown}, {Baliber}, {Bianco}, {Bowman},
  {Burleson}, {Conway}, {Crellin}, {Depagne}, {De Vera}, {Dilday}, {Dragomir},
  {Dubberley}, {Eastman}, {Elphick}, {Falarski}, {Foale}, {Ford}, {Fulton},
  {Garza}, {Gomez}, {Graham}, {Greene}, {Haldeman}, {Hawkins}, {Haworth},
  {Haynes}, {Hidas}, {Hjelstrom}, {Howell}, {Hygelund}, {Lister}, {Lobdill},
  {Martinez}, {Mullins}, {Norbury}, {Parrent}, {Paulson}, {Petry}, {Pickles},
  {Posner}, {Rosing}, {Ross}, {Sand}, {Saunders}, {Shobbrook}, {Shporer},
  {Street}, {Thomas}, {Tsapras}, {Tufts}, {Valenti}, {Vander Horst}, {Walker},
  {White}, \& {Willis}}]{brown13}
{Brown}, T.~M., {et~al.} 2013, \pasp, 125, 1031

\bibitem[{{Buchhave} {et~al.}(2012){Buchhave}, {Latham}, {Johansen},
  {Bizzarro}, {Torres}, {Rowe}, {Batalha}, {Borucki}, {Brugamyer}, {Caldwell},
  {Bryson}, {Ciardi}, {Cochran}, {Endl}, {Esquerdo}, {Ford}, {Geary},
  {Gilliland}, {Hansen}, {Isaacson}, {Laird}, {Lucas}, {Marcy}, {Morse},
  {Robertson}, {Shporer}, {Stefanik}, {Still}, \& {Quinn}}]{buchhave12}
{Buchhave}, L.~A., {et~al.} 2012, \nat, 486, 375

\bibitem[{{Burke} {et~al.}(2015){Burke}, {Christiansen}, {Mullally}, {Seader},
  {Huber}, {Rowe}, {Coughlin}, {Thompson}, {Catanzarite}, {Clarke}, {Morton},
  {Caldwell}, {Bryson}, {Haas}, {Batalha}, {Jenkins}, {Tenenbaum}, {Twicken},
  {Li}, {Quintana}, {Barclay}, {Henze}, {Borucki}, {Howell}, \&
  {Still}}]{burke15}
{Burke}, C.~J., {et~al.} 2015, \apj, 809, 8

\bibitem[{{Campante} {et~al.}(2015){Campante}, {Barclay}, {Swift}, {Huber},
  {Adibekyan}, {Cochran}, {Burke}, {Isaacson}, {Quintana}, {Davies}, {Silva
  Aguirre}, {Ragozzine}, {Riddle}, {Baranec}, {Basu}, {Chaplin},
  {Christensen-Dalsgaard}, {Metcalfe}, {Bedding}, {Handberg}, {Stello},
  {Brewer}, {Hekker}, {Karoff}, {Kolbl}, {Law}, {Lundkvist}, {Miglio}, {Rowe},
  {Santos}, {Van Laerhoven}, {Arentoft}, {Elsworth}, {Fischer}, {Kawaler},
  {Kjeldsen}, {Lund}, {Marcy}, {Sousa}, {Sozzetti}, \& {White}}]{campante15}
{Campante}, T.~L., {et~al.} 2015, \apj, 799, 170

\bibitem[{{Carter} {et~al.}(2012){Carter}, {Agol}, {Chaplin}, {Basu},
  {Bedding}, {Buchhave}, {Christensen-Dalsgaard}, {Deck}, {Elsworth},
  {Fabrycky}, {Ford}, {Fortney}, {Hale}, {Handberg}, {Hekker}, {Holman},
  {Huber}, {Karoff}, {Kawaler}, {Kjeldsen}, {Lissauer}, {Lopez}, {Lund},
  {Lundkvist}, {Metcalfe}, {Miglio}, {Rogers}, {Stello}, {Borucki}, {Bryson},
  {Christiansen}, {Cochran}, {Geary}, {Gilliland}, {Haas}, {Hall}, {Howard},
  {Jenkins}, {Klaus}, {Koch}, {Latham}, {MacQueen}, {Sasselov}, {Steffen},
  {Twicken}, \& {Winn}}]{carter12}
{Carter}, J.~A., {et~al.} 2012, Science, 337, 556

\bibitem[{{Chabrier} \& {Baraffe}(2007)}]{chabrier07}
{Chabrier}, G., \& {Baraffe}, I. 2007, \apjl, 661, L81

\bibitem[{{Chaplin} \& {Miglio}(2013)}]{chaplin13d}
{Chaplin}, W.~J., \& {Miglio}, A. 2013, \araa, 51, 353

\bibitem[{{Chaplin} {et~al.}(2013){Chaplin}, {Sanchis-Ojeda}, {Campante},
  {Handberg}, {Stello}, {Winn}, {Basu}, {Christensen-Dalsgaard}, {Davies},
  {Metcalfe}, {Buchhave}, {Fischer}, {Bedding}, {Cochran}, {Elsworth},
  {Gilliland}, {Hekker}, {Huber}, {Isaacson}, {Karoff}, {Kawaler}, {Kjeldsen},
  {Latham}, {Lund}, {Lundkvist}, {Marcy}, {Miglio}, {Barclay}, \&
  {Lissauer}}]{chaplin13c}
{Chaplin}, W.~J., {et~al.} 2013, \apj, 766, 101

\bibitem[{{Christensen-Dalsgaard} {et~al.}(2010){Christensen-Dalsgaard},
  {Kjeldsen}, {Brown}, {Gilliland}, {Arentoft}, {Frandsen}, {Quirion},
  {Borucki}, {Koch}, \& {Jenkins}}]{cd10}
{Christensen-Dalsgaard}, J., {et~al.} 2010, \apjl, 713, L164

\bibitem[{{Ciceri} {et~al.}(2015){Ciceri}, {Lillo-Box}, {Southworth},
  {Mancini}, {Henning}, \& {Barrado}}]{ciceri15}
{Ciceri}, S., et al. 2015, \aap, 573, L5

\bibitem[{{Davies} {et~al.}(2015){Davies}, {Silva Aguirre}, {Bedding},
  {Handberg}, {Lund}, {Chaplin}, {Huber}, {White}, {Benomar}, {Hekker}, {Basu},
  {Campante}, {Christensen-Dalsgaard}, {Elsworth}, {Karoff}, {Kjeldsen},
  {Lundkvist}, {Metcalfe}, \& {Stello}}]{davies15}
{Davies}, G.~R., {et~al.} 2015, \mnras, in press (arXiv:1511.02105)

\bibitem[{{Dong} \& {Zhu}(2013)}]{dong13b}
{Dong}, S., \& {Zhu}, Z. 2013, \apj, 778, 53

\bibitem[{{Gilliland} {et~al.}(2011){Gilliland}, {McCullough}, {Nelan},
  {Brown}, {Charbonneau}, {Nutzman}, {Christensen-Dalsgaard}, \&
  {Kjeldsen}}]{gilliland11}
{Gilliland}, R.~L., et al. 2011, \apj, 726, 2

\bibitem[{{Gizon} \& {Solanki}(2003)}]{gizon03}
{Gizon}, L., \& {Solanki}, S.~K. 2003, \apj, 589, 1009

\bibitem[{{Gizon} {et~al.}(2013){Gizon}, {Ballot}, {Michel}, {Stahn},
  {Vauclair}, {Bruntt}, {Quirion}, {Benomar}, {Vauclair}, {Appourchaux},
  {Auvergne}, {Baglin}, {Barban}, {Baudin}, {Bazot}, {Campante}, {Catala},
  {Chaplin}, {Creevey}, {Deheuvels}, {Dolez}, {Elsworth}, {Garcia}, {Gaulme},
  {Mathis}, {Mathur}, {Mosser}, {Regulo}, {Roxburgh}, {Salabert}, {Samadi},
  {Sato}, {Verner}, {Hanasoge}, \& {Sreenivasan}}]{gizon13}
{Gizon}, L., {et~al.} 2013, Proceedings of the National Academy of Science,
  110, 13267

\bibitem[{{Grundahl} {et~al.}(2008){Grundahl}, {Christensen-Dalsgaard},
  {Kjeldsen}, {Frandsen}, {Arentoft}, {Kjaergaard}, \&
  {J{\o}rgensen}}]{grundahl08}
{Grundahl}, F. et al. 2008, in IAU
  Symposium, Vol. 252, IAU Symposium, ed. {L.~Deng \& K.~L.~Chan}, 465--466

\bibitem[{{Howard} {et~al.}(2012){Howard}, {Marcy}, {Bryson}, {Jenkins},
  {Rowe}, {Batalha}, {Borucki}, {Koch}, {Dunham}, {Gautier}, {Van Cleve},
  {Cochran}, {Latham}, {Lissauer}, {Torres}, {Brown}, {Gilliland}, {Buchhave},
  {Caldwell}, {Christensen-Dalsgaard}, {Ciardi}, {Fressin}, {Haas}, {Howell},
  {Kjeldsen}, {Seager}, {Rogers}, {Sasselov}, {Steffen}, {Basri},
  {Charbonneau}, {Christiansen}, {Clarke}, {Dupree}, {Fabrycky}, {Fischer},
  {Ford}, {Fortney}, {Tarter}, {Girouard}, {Holman}, {Johnson}, {Klaus},
  {Machalek}, {Moorhead}, {Morehead}, {Ragozzine}, {Tenenbaum}, {Twicken},
  {Quinn}, {Isaacson}, {Shporer}, {Lucas}, {Walkowicz}, {Welsh}, {Boss},
  {Devore}, {Gould}, {Smith}, {Morris}, {Prsa}, {Morton}, {Still}, {Thompson},
  {Mullally}, {Endl}, \& {MacQueen}}]{howard11}
{Howard}, A.~W., {et~al.} 2012, \apjs, 201, 15

\bibitem[{{Howell} {et~al.}(2011){Howell}, {Everett}, {Sherry}, {Horch}, \&
  {Ciardi}}]{howell11}
{Howell}, S.~B., {Everett}, M.~E., {Sherry}, W., {Horch}, E., \& {Ciardi},
  D.~R. 2011, \aj, 142, 19

\bibitem[{{Howell} {et~al.}(2014){Howell}, {Sobeck}, {Haas}, {Still},
  {Barclay}, {Mullally}, {Troeltzsch}, {Aigrain}, {Bryson}, {Caldwell},
  {Chaplin}, {Cochran}, {Huber}, {Marcy}, {Miglio}, {Najita}, {Smith},
  {Twicken}, \& {Fortney}}]{howell14}
{Howell}, S.~B., {et~al.} 2014, \pasp, 126, 398

\bibitem[{{Huber}(2015)}]{huber14b}
{Huber}, D. 2015, in Astrophysics and Space Science Library, Vol. 408, Giants
  of Eclipse: The zeta Aurigae Stars and Other Binary Systems, 169

\bibitem[{{Huber} {et~al.}(2013{\natexlab{a}}){Huber}, {Chaplin},
  {Christensen-Dalsgaard}, {Gilliland}, {Kjeldsen}, {Buchhave}, {Fischer},
  {Lissauer}, {Rowe}, {Sanchis-Ojeda}, {Basu}, {Handberg}, {Hekker}, {Howard},
  {Isaacson}, {Karoff}, {Latham}, {Lund}, {Lundkvist}, {Marcy}, {Miglio},
  {Silva Aguirre}, {Stello}, {Arentoft}, {Barclay}, {Bedding}, {Burke},
  {Christiansen}, {Elsworth}, {Haas}, {Kawaler}, {Metcalfe}, {Mullally}, \&
  {Thompson}}]{huber13}
{Huber}, D., {et~al.} 2013{\natexlab{a}}, \apj, 767, 127

\bibitem[{{Huber} {et~al.}(2013{\natexlab{b}}){Huber}, {Carter}, {Barbieri},
  {Miglio}, {Deck}, {Fabrycky}, {Montet}, {Buchhave}, {Chaplin}, {Hekker},
  {Montalb{\'a}n}, {Sanchis-Ojeda}, {Basu}, {Bedding}, {Campante},
  {Christensen-Dalsgaard}, {Elsworth}, {Stello}, {Arentoft}, {Ford},
  {Gilliland}, {Handberg}, {Howard}, {Isaacson}, {Johnson}, {Karoff},
  {Kawaler}, {Kjeldsen}, {Latham}, {Lund}, {Lundkvist}, {Marcy}, {Metcalfe},
  {Silva Aguirre}, \& {Winn}}]{huber13b}
---. 2013{\natexlab{b}}, Science, 342, 331

\bibitem[{{Jenkins} {et~al.}(2015){Jenkins}, {Twicken}, {Batalha}, {Caldwell},
  {Cochran}, {Endl}, {Latham}, {Esquerdo}, {Seader}, {Bieryla}, {Petigura},
  {Ciardi}, {Marcy}, {Isaacson}, {Huber}, {Rowe}, {Torres}, {Bryson},
  {Buchhave}, {Ramirez}, {Wolfgang}, {Li}, {Campbell}, {Tenenbaum},
  {Sanderfer}, {Henze}, {Catanzarite}, {Gilliland}, \& {Borucki}}]{jenkins15}
{Jenkins}, J.~M., {et~al.} 2015, \aj, 150, 56

\bibitem[{{Johnson} {et~al.}(2013){Johnson}, {Morton}, \& {Wright}}]{johnson13}
{Johnson}, J.~A., {Morton}, T.~D., \& {Wright}, J.~T. 2013, \apj, 763, 53

\bibitem[{{Kipping}(2010)}]{kipping10}
{Kipping}, D.~M. 2010, \mnras, 407, 301

\bibitem[{{Kjeldsen} \& {Bedding}(1995)}]{kb95}
{Kjeldsen}, H., \& {Bedding}, T.~R. 1995, \aap, 293, 87

\bibitem[{{Lebreton} \& {Goupil}(2014)}]{lebreton14}
{Lebreton}, Y., \& {Goupil}, M.~J. 2014, \aap, 569, A21

\bibitem[{{Lillo-Box} {et~al.}(2014){Lillo-Box}, {Barrado}, {Henning},
  {Mancini}, {Ciceri}, {Figueira}, {Santos}, {Aceituno}, \&
  {S{\'a}nchez}}]{lillobox14}
{Lillo-Box}, J., {et~al.} 2014, \aap, 568, L1

\bibitem[{{Lissauer} {et~al.}(2011){Lissauer}, {Fabrycky}, {Ford}, {Borucki},
  {Fressin}, {Marcy}, {Orosz}, {Rowe}, {Torres}, {Welsh}, {Batalha}, {Bryson},
  {Buchhave}, {Caldwell}, {Carter}, {Charbonneau}, {Christiansen}, {Cochran},
  {Desert}, {Dunham}, {Fanelli}, {Fortney}, {Gautier}, {Geary}, {Gilliland},
  {Haas}, {Hall}, {Holman}, {Koch}, {Latham}, {Lopez}, {McCauliff}, {Miller},
  {Morehead}, {Quintana}, {Ragozzine}, {Sasselov}, {Short}, \&
  {Steffen}}]{lissauer11}
{Lissauer}, J.~J., {et~al.} 2011, \nat, 470, 53

\bibitem[{{Lloyd}(2011)}]{lloyd11}
{Lloyd}, J.~P. 2011, \apjl, 739, L49

\bibitem[{{Lopez} \& {Fortney}(2015)}]{lopez15}
{Lopez}, E.~D., \& {Fortney}, J.~J. 2015, \apj, in press (arXiv:1510.00067)

\bibitem[{{Lund} {et~al.}(2014){Lund}, {Lundkvist}, {Silva Aguirre}, {Houdek},
  {Casagrande}, {Van Eylen}, {Campante}, {Karoff}, {Kjeldsen}, {Albrecht},
  {Chaplin}, {Nielsen}, {Degroote}, {Davies}, \& {Handberg}}]{lund14}
{Lund}, M.~N., {et~al.} 2014, \aap, 570, A54

\bibitem[{{Metcalfe} {et~al.}(2012){Metcalfe}, {Chaplin}, {Appourchaux},
  {Garc{\'{\i}}a}, {Basu}, {Brand{\~a}o}, {Creevey}, {Deheuvels}, {Do{\u g}an},
  {Eggenberger}, {Karoff}, {Miglio}, {Stello}, {Y{\i}ld{\i}z}, {{\c C}elik},
  {Antia}, {Benomar}, {Howe}, {R{\'e}gulo}, {Salabert}, {Stahn}, {Bedding},
  {Davies}, {Elsworth}, {Gizon}, {Hekker}, {Mathur}, {Mosser}, {Bryson},
  {Still}, {Christensen-Dalsgaard}, {Gilliland}, {Kawaler}, {Kjeldsen},
  {Ibrahim}, {Klaus}, \& {Li}}]{metcalfe12}
{Metcalfe}, T.~S., {et~al.} 2012, \apjl, 748, L10

\bibitem[{{Mullally} {et~al.}(2015){Mullally}, {Coughlin}, {Thompson}, {Rowe},
  {Burke}, {Latham}, {Batalha}, {Bryson}, {Christiansen}, {Henze}, {Ofir},
  {Quarles}, {Shporer}, {Van Eylen}, {Van Laerhoven}, {Shah}, {Wolfgang},
  {Chaplin}, {Xie}, {Akeson}, {Argabright}, {Bachtell}, {Barclay}, {Borucki},
  {Caldwell}, {Campbell}, {Catanzarite}, {Cochran}, {Duren}, {Fleming},
  {Fraquelli}, {Girouard}, {Haas}, {He{\l}miniak}, {Howell}, {Huber}, {Larson},
  {Gautier}, {Jenkins}, {Li}, {Lissauer}, {McArthur}, {Miller}, {Morris},
  {Patil-Sabale}, {Plavchan}, {Putnam}, {Quintana}, {Ramirez}, {Silva Aguirre},
  {Seader}, {Smith}, {Steffen}, {Stewart}, {Stober}, {Still}, {Tenenbaum},
  {Troeltzsch}, {Twicken}, \& {Zamudio}}]{mullally15}
{Mullally}, F., {et~al.} 2015, \apjs, 217, 31

\bibitem[{{Nutzman} {et~al.}(2011){Nutzman}, {Gilliland}, {McCullough},
  {Charbonneau}, {Christensen-Dalsgaard}, {Kjeldsen}, {Nelan}, {Brown}, \&
  {Holman}}]{nutzman11}
{Nutzman}, P., {et~al.} 2011, \apj, 726, 3

\bibitem[{{Petigura} {et~al.}(2013){Petigura}, {{\SortNoop{z}}Howard}, \&
  {Marcy}}]{petigura13b}
{Petigura}, E.~A., {{\SortNoop{z}}Howard}, A.~W., \& {Marcy}, G.~W. 2013, PNAS,
  110, 19175

\bibitem[{{Quinn} {et~al.}(2015){Quinn}, {White}, {Latham}, {Chaplin},
  {Handberg}, {Huber}, {Kipping}, {Payne}, {Jiang}, {Silva Aguirre}, {Stello},
  {Sliski}, {Ciardi}, {Buchhave}, {Bedding}, {Davies}, {Hekker}, {Kjeldsen},
  {Kuszlewicz}, {Everett}, {Howell}, {Basu}, {Campante},
  {Christensen-Dalsgaard}, {Elsworth}, {Karoff}, {Kawaler}, {Lund},
  {Lundkvist}, {Esquerdo}, {Calkins}, \& {Berlind}}]{quinn15}
{Quinn}, S.~N., {et~al.} 2015, \apj, 803, 49

\bibitem[{{Quintana} {et~al.}(2014){Quintana}, {Barclay}, {Raymond}, {Rowe},
  {Bolmont}, {Caldwell}, {Howell}, {Kane}, {Huber}, {Crepp}, {Lissauer},
  {Ciardi}, {Coughlin}, {Everett}, {Henze}, {Horch}, {Isaacson}, {Ford},
  {Adams}, {Still}, {Hunter}, {Quarles}, \& {Selsis}}]{quintana14}
{Quintana}, E.~V., {et~al.} 2014, Science, 344, 277

\bibitem[{{Rauer} {et~al.}(2014){Rauer}, {Catala}, {Aerts}, {Appourchaux},
  {Benz}, {Brandeker}, {Christensen-Dalsgaard}, {Deleuil}, {Gizon}, {Goupil},
  {G{\"u}del}, {Janot-Pacheco}, {Mas-Hesse}, {Pagano}, {Piotto}, {Pollacco},
  {Santos}, {Smith}, {Su{\'a}rez}, {Szab{\'o}}, {Udry}, {Adibekyan}, {Alibert},
  {Almenara}, {Amaro-Seoane}, {Ammer-von Eiff}, {Asplund}, {Antonello},
  {Barnes}, {Baudin}, {Belkacem}, {Bergemann}, {Bihain}, {Birch}, {Bonfils},
  {Boisse}, {Bonomo}, {Borsa}, {Brand{\~a}o}, {Brocato}, {Brun}, {Burleigh},
  {Burston}, {Cabrera}, {Cassisi}, {Chaplin}, {Charpinet}, {Chiappini},
  {Church}, {Csizmadia}, {Cunha}, {Damasso}, {Davies}, {Deeg}, {D{\'{\i}}az},
  {Dreizler}, {Dreyer}, {Eggenberger}, {Ehrenreich}, {Eigm{\"u}ller},
  {Erikson}, {Farmer}, {Feltzing}, {Oliveira Fialho}, {Figueira}, {Forveille},
  {Fridlund}, {Garc{\'{\i}}a}, {Giommi}, {Giuffrida}, {Godolt}, {Gomes da
  Silva}, {Granzer}, {Grenfell}, {Grotsch-Noels}, {G{\"u}nther}, {Haswell},
  {Hatzes}, {H{\'e}brard}, {Hekker}, {Helled}, {Heng}, {Jenkins}, {Johansen},
  {Khodachenko}, {Kislyakova}, {Kley}, {Kolb}, {Krivova}, {Kupka}, {Lammer},
  {Lanza}, {Lebreton}, {Magrin}, {Marcos-Arenal}, {Marrese}, {Marques},
  {Martins}, {Mathis}, {Mathur}, {Messina}, {Miglio}, {Montalban}, {Montalto},
  {Monteiro}, {Moradi}, {Moravveji}, {Mordasini}, {Morel}, {Mortier},
  {Nascimbeni}, {Nelson}, {Nielsen}, {Noack}, {Norton}, {Ofir}, {Oshagh},
  {Ouazzani}, {P{\'a}pics}, {Parro}, {Petit}, {Plez}, {Poretti}, {Quirrenbach},
  {Ragazzoni}, {Raimondo}, {Rainer}, {Reese}, {Redmer}, {Reffert},
  {Rojas-Ayala}, {Roxburgh}, {Salmon}, {Santerne}, {Schneider}, {Schou},
  {Schuh}, {Schunker}, {Silva-Valio}, {Silvotti}, {Skillen}, {Snellen}, {Sohl},
  {Sousa}, {Sozzetti}, {Stello}, {Strassmeier}, {{\v S}vanda}, {Szab{\'o}},
  {Tkachenko}, {Valencia}, {Van Grootel}, {Vauclair}, {Ventura}, {Wagner},
  {Walton}, {Weingrill}, {Werner}, {Wheatley}, \& {Zwintz}}]{rauer14}
{Rauer}, H., {et~al.} 2014, Experimental Astronomy

\bibitem[{{Ricker} {et~al.}(2014){Ricker}, {Winn}, {Vanderspek}, {Latham},
  {Bakos}, {Bean}, {Berta-Thompson}, {Brown}, {Buchhave}, {Butler}, {Butler},
  {Chaplin}, {Charbonneau}, {Christensen-Dalsgaard}, {Clampin}, {Deming},
  {Doty}, {De Lee}, {Dressing}, {Dunham}, {Endl}, {Fressin}, {Ge}, {Henning},
  {Holman}, {Howard}, {Ida}, {Jenkins}, {Jernigan}, {Johnson}, {Kaltenegger},
  {Kawai}, {Kjeldsen}, {Laughlin}, {Levine}, {Lin}, {Lissauer}, {MacQueen},
  {Marcy}, {McCullough}, {Morton}, {Narita}, {Paegert}, {Palle}, {Pepe},
  {Pepper}, {Quirrenbach}, {Rinehart}, {Sasselov}, {Sato}, {Seager},
  {Sozzetti}, {Stassun}, {Sullivan}, {Szentgyorgyi}, {Torres}, {Udry}, \&
  {Villasenor}}]{ricker14}
{Ricker}, G.~R., {et~al.} 2014, in Society of Photo-Optical Instrumentation
  Engineers (SPIE) Conference Series, Vol. 9143, , 20

\bibitem[{{Sanchis-Ojeda} {et~al.}(2013){Sanchis-Ojeda}, {Winn}, {Marcy},
  {Howard}, {Isaacson}, {Johnson}, {Torres}, {Albrecht}, {Campante}, {Chaplin},
  {Davies}, {Lund}, {Carter}, {Dawson}, {Buchhave}, {Everett}, {Fischer},
  {Geary}, {Gilliland}, {Horch}, {Howell}, \& {Latham}}]{sanchis13}
{Sanchis-Ojeda}, R., {et~al.} 2013, \apj, 775, 54

\bibitem[{{Schlaufman} \& {Winn}(2013)}]{schlaufman13}
{Schlaufman}, K.~C., \& {Winn}, J.~N. 2013, \apj, 772, 143

\bibitem[{{Seager} \& {Mall{\'e}n-Ornelas}(2003)}]{seager03}
{Seager}, S., \& {Mall{\'e}n-Ornelas}, G. 2003, \apj, 585, 1038

\bibitem[{{Silva Aguirre} {et~al.}(2015){Silva Aguirre}, {Davies}, {Basu},
  {Christensen-Dalsgaard}, {Creevey}, {Metcalfe}, {Bedding}, {Casagrande},
  {Handberg}, {Lund}, {Nissen}, {Chaplin}, {Huber}, {Serenelli}, {Stello}, {Van
  Eylen}, {Campante}, {Elsworth}, {Gilliland}, {Hekker}, {Karoff}, {Kawaler},
  {Kjeldsen}, \& {Lundkvist}}]{silva15}
{Silva Aguirre}, V., {et~al.} 2015, \mnras, 452, 2127

\bibitem[{{Sliski} \& {Kipping}(2014)}]{sliski14}
{Sliski}, D.~H., \& {Kipping}, D.~M. 2014, \apj, 788, 148

\bibitem[{{Steffen} {et~al.}(2012){Steffen}, {Fabrycky}, {Ford}, {Carter},
  {D{\'e}sert}, {Fressin}, {Holman}, {Lissauer}, {Moorhead}, {Rowe},
  {Ragozzine}, {Welsh}, {Batalha}, {Borucki}, {Buchhave}, {Bryson}, {Caldwell},
  {Charbonneau}, {Ciardi}, {Cochran}, {Endl}, {Everett}, {Gautier},
  {Gilliland}, {Girouard}, {Jenkins}, {Horch}, {Howell}, {Isaacson}, {Klaus},
  {Koch}, {Latham}, {Li}, {Lucas}, {MacQueen}, {Marcy}, {McCauliff}, {Middour},
  {Morris}, {Mullally}, {Quinn}, {Quintana}, {Shporer}, {Still}, {Tenenbaum},
  {Thompson}, {Twicken}, \& {Van Cleve}}]{steffen12}
{Steffen}, J.~H., {et~al.} 2012, \mnras, 421, 2342

\bibitem[{{Stello} {et~al.}(2009){Stello}, {Chaplin}, {Bruntt}, {Creevey},
  {Garc{\'{\i}}a-Hern{\'a}ndez}, {Monteiro}, {Moya}, {Quirion}, {Sousa},
  {Su{\'a}rez}, {Appourchaux}, {Arentoft}, {Ballot}, {Bedding},
  {Christensen-Dalsgaard}, {Elsworth}, {Fletcher}, {Garc{\'{\i}}a}, {Houdek},
  {Jim{\'e}nez-Reyes}, {Kjeldsen}, {New}, {R{\'e}gulo}, {Salabert}, \&
  {Toutain}}]{stello09}
{Stello}, D., {et~al.} 2009, \apj, 700, 1589

\bibitem[{{Sullivan} {et~al.}(2015){Sullivan}, {Winn}, {Berta-Thompson},
  {Charbonneau}, {Deming}, {Dressing}, {Latham}, {Levine}, {McCullough},
  {Morton}, {Ricker}, {Vanderspek}, \& {Woods}}]{sullivan15}
{Sullivan}, P.~W., {et~al.} 2015, \apj, 809, 77

\bibitem[{{Van Eylen} \& {Albrecht}(2015)}]{vaneylen15}
{Van Eylen}, V., \& {Albrecht}, S. 2015, \apj, 808, 126

\bibitem[{{Van Eylen} {et~al.}(2014){Van Eylen}, {Lund}, {Silva Aguirre},
  {Arentoft}, {Kjeldsen}, {Albrecht}, {Chaplin}, {Isaacson}, {Pedersen},
  {Jessen-Hansen}, {Tingley}, {Christensen-Dalsgaard}, {Aerts}, {Campante}, \&
  {Bryson}}]{vaneylen14}
{Van Eylen}, V., {et~al.} 2014, \apj, 782, 14

\bibitem[{{White} {et~al.}(2011){White}, {Bedding}, {Stello},
  {Christensen-Dalsgaard}, {Huber}, \& {Kjeldsen}}]{white11}
{White}, et al. 2011, \apj, 743, 161

\bibitem[{{Winn} {et~al.}(2010){Winn}, {Fabrycky}, {Albrecht}, \&
  {Johnson}}]{winn10}
{Winn}, J.~N., {Fabrycky}, D., {Albrecht}, S., \& {Johnson}, J.~A. 2010, \apjl,
  718, L145

\end{thebibliography}

\end{document}